\documentclass[prb,twocolumn,lengthcheck]{revtex4}%
\usepackage{amsfonts}
\usepackage{amsmath}
\usepackage{amssymb}
\usepackage{graphicx}%
\ifx\pdfoutput\relax\let\pdfoutput=\undefined\fi
\newcount\msipdfoutput
\ifx\pdfoutput\undefined\else
\ifcase\pdfoutput\else
\ifx\paperwidth\undefined\else
\ifdim\paperheight=0pt\relax\else\pdfpageheight\paperheight\fi
\ifdim\paperwidth=0pt\relax\else\pdfpagewidth\paperwidth\fi
\fi\fi\fi
\begin{document}
\title{Quantum dynamical phase transition in a system with many-body interactions}
\author{E. P. Danieli}
\affiliation{Facultad de Matem\'{a}tica, Astronom\'{\i}a y F\'{\i}sica, Universidad
Nacional de C\'{o}rdoba, 5000 C\'{o}rdoba, Argentina}
\author{G. A. \'{A}lvarez}
\affiliation{Facultad de Matem\'{a}tica, Astronom\'{\i}a y F\'{\i}sica, Universidad
Nacional de C\'{o}rdoba, 5000 C\'{o}rdoba, Argentina}
\author{P. R. Levstein}
\affiliation{Facultad de Matem\'{a}tica, Astronom\'{\i}a y F\'{\i}sica, Universidad
Nacional de C\'{o}rdoba, 5000 C\'{o}rdoba, Argentina}
\author{H. M. Pastawski}
\affiliation{Facultad de Matem\'{a}tica, Astronom\'{\i}a y F\'{\i}sica, Universidad
Nacional de C\'{o}rdoba, 5000 C\'{o}rdoba, Argentina}

\begin{abstract}
We introduce a microscopic Hamiltonian model of a two level system \ with
many-body interactions with an environment whose excitation dynamics is fully
solved within the Keldysh formalism. If a particle starts in one of the states
of the isolated system, the return probability oscillates with the Rabi
frequency $\omega_{0}$. For weak interactions with the environment
$1/\tau_{\mathrm{SE}}^{{}}<2\omega_{0},$ we find a slower oscillation whose
amplitude decays with a rate $1/\tau_{\phi}=1/(2\tau_{\mathrm{SE}}^{{}}).$
However, beyond a finite critical interaction with the environment,
$1/\tau_{\mathrm{SE}}^{{}}>2\omega_{0},$ the decay rate becomes $1/\tau_{\phi
}^{{}}\propto\omega_{0}^{2}~\tau_{\mathrm{SE}}^{{}}$. The oscillation period
diverges showing a \emph{quantum dynamical phase transition }to a Quantum Zeno phase.

\end{abstract}
\volumeyear{year}
\volumenumber{number}
\issuenumber{number}
\eid{identifier}
\date[Date text]{date}
\received[Received text]{date}

\revised[Revised text]{date}

\accepted[Accepted text]{date}

\published[Published text]{date}

\maketitle

Ideal quantum information processing (QIP) involves manipulating a system's
Hamiltonian. In practice, the interactions with an environment
\cite{Zurek2003} perturb the evolution, smoothly degrading the quantum
interferences within a \textquotedblleft decoherence\textquotedblright\ rate,
$1/\tau_{\phi}$. Although one expects $1/\tau_{\phi}$ to be proportional to
the system-environment (SE) interaction rate $1/\tau_{\mathrm{SE}}$,\ there
are conditions where $1/\tau_{\phi}$\ does not dependent on it \cite{PhysicaA}%
. This phenomenon was interpreted \cite{JalPas,Beenakker,CookPasJal} as the
onset of a Lyapunov phase, where the decay rate is the Lyapunov exponent
$\lambda$ characterizing the complexity of the classical system. The
description of this transition, $1/\tau_{\phi}=\min\left[  1/\tau
_{\mathrm{SE}},\lambda\right]  $, requires evaluation of the observables
beyond perturbation theory \cite{JalPas,CookPasJal}. We will show that a
dynamical transition also occurs in a swapping gate: a system that jumps
between two equivalent states, $A$ and $B$, when the coupling $V_{AB}$ is
turned on \cite{exp-swapp, Nakamura}. Starting on state $A$, the return
probability oscillates with the Rabi frequency $\omega_{0}=2V_{AB}/\hslash$.
Since each state interacts with the environment at a rate $1/\tau
_{\mathrm{SE}}^{{}}$, weak interactions ($1/\tau_{\mathrm{SE}}^{{}}%
<2\omega_{0}$) produce a slightly slower oscillation which decays at a rate
$1/\tau_{\phi}=1/(2\tau_{\mathrm{SE}}^{{}})$. However, the swapping frequency
is\textit{ non-analytic} on the interaction rate and at a critical strength
$1/\tau_{\mathrm{SE}}^{c}=2\omega_{0}$, the oscillation freezes indicating a
\textit{transition} to a new\textit{ dynamical regime}. The initial state now
decays to equilibrium at a slow rate $1/\tau_{\phi}^{{}}\propto\omega_{0}%
^{2}~\tau_{\mathrm{SE}}^{{}}$ which cancels for strong SE interaction. This
last regime can be seen as a Quantum Zeno phase, where the dynamics is
inhibited by frequent \textquotedblleft observations\textquotedblright%
\ \cite{Misra-Sudarshan} by the environment. Such quantum freeze can arise as
a pure dynamical process governed by strictly unitary evolutions
\cite{Pascazio} \cite{Pastawski-Usaj}.

\begin{figure}
[h]
\begin{center}
\includegraphics[
height=2.7256in,
width=2.9772in
]%
{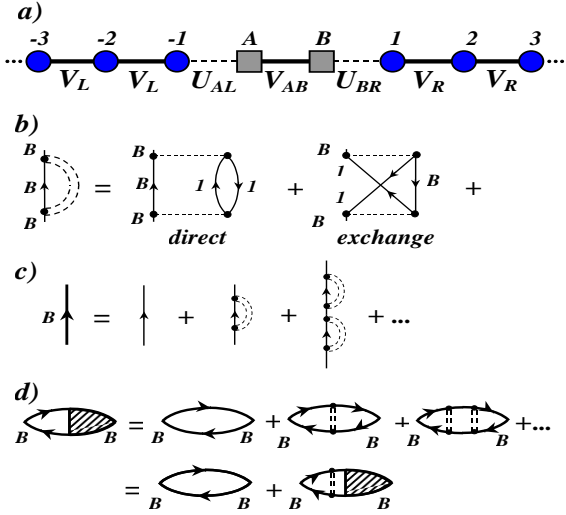}%
\caption{(Color online) a) System-environment scheme. Squares and circles
represent system and environment states respectively. Solid lines are
tunneling matrix elements while dashed lines are through-space Coulomb
interactions. b) The relevant self-enegy diagrams in a local basis. Lines are
exact Green's functions of the system and the environment in absence of
Coulomb interactions (dashed lines). c) Retarded Green's function of the
system-environment at site $B$. It contains the Coulomb interaction through
the self-energy correction of b) to infinite order. d) Particle density
function at site $B$. The dashed lines represent local interactions in time
and space.}%
\label{Fig_feynmann_ising}%
\end{center}
\end{figure}

We consider a \textquotedblleft system\textquotedblright\ with a
\textit{single} electron occupying one of two coupled states, $A$ or $B,$ each
interacting with a corresponding electron reservoir (the \textquotedblleft
environment\textquotedblright). The total system, represented in Fig.
\ref{Fig_feynmann_ising} a) has the Hamiltonian $\widehat{\mathcal{H}%
}=\widehat{\mathcal{H}}_{\mathrm{S}}+\widehat{\mathcal{H}}_{\mathrm{E}%
}+\widehat{\mathcal{H}}_{\mathrm{SE}}$, where the first term is
\begin{equation}
\widehat{\mathcal{H}}_{\mathrm{S}}=E_{A}^{{}}\hat{c}_{A}^{+}\hat{c}_{A}^{{}%
}+E_{B}^{{}}\hat{c}_{B}^{+}\hat{c}_{B}^{{}}-V_{AB}\left(  \hat{c}_{A}^{+}%
\hat{c}_{B}^{{}}+\hat{c}_{B}^{+}\hat{c}_{A}^{{}}\right)  . \label{HS}%
\end{equation}
Here, $\hat{c}_{i}^{+}(\hat{c}_{i}^{{}})$ are the standard fermionic creation
(destruction) operators. $E_{i}$ stands for the energy of the $i$-th local
state whose spin index is omitted. $V_{AB}$ yields the natural frequency,
$\omega_{0}=2V_{AB}/\hbar$. The environment is
\begin{equation}
\widehat{\mathcal{H}}_{\mathrm{E}}=\sum_{\substack{i=-\infty\\(i\neq
0)}}^{\infty}\left(  E_{i/\left\vert i\right\vert }^{{}}\hat{c}_{i}^{+}\hat
{c}_{i}^{{}}-V_{i/\left\vert i\right\vert }^{{}}\left(  \hat{c}_{i}^{+}\hat
{c}_{i+i/\left\vert i\right\vert }^{{}}+\hat{c}_{i+i/\left\vert i\right\vert
}^{+}\hat{c}_{i}^{{}}\right)  \right)
\end{equation}
where the sums on negative (positive) index describe a semi-infinite chain to
the left \ (right) acting as a reservoir. $E_{-1}\equiv E_{L}$ \ and
$E_{1}\equiv E_{R}$ are site energies while $V_{-1}\equiv V_{L}$ and
$V_{1}\equiv V_{R}$ are adjacent site hoppings.\ The system-environment
interaction is modeled with a through-space interaction
\begin{gather}
\widehat{\mathcal{H}}_{\mathrm{SE}}=\sum_{\alpha=\uparrow,\downarrow}\left\{
\sum_{\beta=\uparrow,\downarrow}U_{B\mathrm{R}}^{(\mathrm{dir.})}~\hat
{c}_{B\beta}^{+}\hat{c}_{B\beta}^{{}}\hat{c}_{1\alpha}^{+}\hat{c}_{1\alpha
}^{{}}\right.  +\label{HSE}\\
\left.  U_{B\mathrm{R}}^{(\mathrm{exch.})}~\hat{c}_{B\alpha}^{+}\hat
{c}_{1\alpha}^{{}}\hat{c}_{1\alpha}^{+}\hat{c}_{B\alpha}^{{}}+\sum
_{\beta=\uparrow,\downarrow}U_{A\mathrm{L}}^{(\mathrm{dir.})}~\hat{c}_{A\beta
}^{+}\hat{c}_{A\beta}^{{}}\hat{c}_{-1\alpha}^{+}\hat{c}_{-1\alpha}^{{}}\right.
\nonumber\\
+\left.  U_{A\mathrm{L}}^{(\mathrm{exch.})}~\hat{c}_{A\alpha}^{+}\hat
{c}_{-1\alpha}^{{}}\hat{c}_{-1\alpha}^{+}\hat{c}_{A\alpha}^{{}}\right\}
.\nonumber
\end{gather}
The first line represents the Coulomb interaction of an electron in state $B$
with an electron in the first site of reservoir to the right. $U_{B\mathrm{R}%
}^{(\mathrm{dir.})}$is the standard direct integral and $U_{B\mathrm{R}%
}^{(\mathrm{exch.})}$ is the exchange one. Analogously, the second line is the
interaction with the reservoir to the left.%

A complete norm preserving solution requires the evaluation of the reduced
particle and hole density functions $G_{ij}^{<}\left(  t_{2,}t_{1}\right)
=\frac{\mathrm{i}}{\hbar}\left\langle \Psi\right\vert \hat{c}_{j}^{+}\left(
t_{1}\right)  \hat{c}_{i}^{{}}\left(  t_{2}\right)  \left\vert \Psi
\right\rangle $ and $G_{ij}^{>}\left(  t_{2,}t_{1}\right)  =-\frac{\mathrm{i}%
}{\hbar}\left\langle \Psi\right\vert \hat{c}_{i}^{{}}\left(  t_{2}\right)
\hat{c}_{j}^{+}\left(  t_{1}\right)  \left\vert \Psi\right\rangle $ that
describe temporal and spacial correlations. Here, the creation and destruction
operators are in the Heisenberg representation and $\left\vert \Psi
\right\rangle =\hat{c}_{A}^{+}\hat{c}_{B}^{{}}\left\vert \Psi_{0}\right\rangle
$ is an initial non-equilibrium many-body state built from the non-interacting
equilibrium $\left\vert \Psi_{0}\right\rangle $. The retarded Green's function
$G_{ij}^{\mathrm{R}}\left(  t_{2},t_{1}\right)  =\left[  G_{ji}^{\mathrm{A}%
}\left(  t_{1},t_{2}\right)  \right]  ^{\dagger}=\theta\left(  t_{2}%
,t_{1}\right)  \ [G_{ij}^{>}\left(  t_{2},t_{1}\right)  -G_{ij}^{<}\left(
t_{2},t_{1}\right)  ]$ describes the probability amplitude of finding an
electron at site $i$ after placing it at site $j$ and letting it evolve under
the total Hamiltonian for a time $t_{2}-t_{1}$. By restricting the analysis to
$i,j\in\left\{  A,B\right\}  $, $\mathbf{G}^{<}\left(  t,t\right)  $ is the
single particle 2$\times$2 density matrix and $\mathbf{G}^{\mathrm{R}}\left(
t_{2},t_{1}\right)  $ is an effective evolution operator in this reduced
space. In absence of SE interaction, the Green's function is easily evaluated
in its energy representation $\mathbf{G}^{0\mathrm{R}}\left(  \varepsilon
\right)  =\int\mathbf{G}^{0\mathrm{R}}\left(  t\right)  \exp[\mathrm{i}%
\varepsilon t/\hbar]\mathrm{d}t=[\varepsilon\mathbf{I-H}_{\mathrm{S}}]^{-1}.$
Conversely, the interacting Green's function defines the reduced effective
Hamiltonian and the self-energies $\mathbf{\Sigma}^{\mathrm{R}}(\varepsilon)$
\cite{DAmato}, $\mathbf{H}_{\mathrm{eff.}}(\varepsilon)\equiv\varepsilon
\mathbf{I}-\left[  \mathbf{G}^{\mathrm{R}}\left(  \varepsilon\right)  \right]
^{-1}=\mathbf{H}_{\mathrm{S}}+\mathbf{\Sigma}^{\mathrm{R}}(\varepsilon)$,
where the exact perturbed dynamics is contained in the nonlinear dependence of
the self-energies $\Sigma^{\mathrm{R}}$ on $\varepsilon$. For infinite
reservoirs $\operatorname{Re}\Sigma^{\mathrm{R}}\left(  \varepsilon_{\nu}%
^{o}\right)  $ represents the \textquotedblleft shift\textquotedblright\ of
the system's eigen-energies $\varepsilon_{\nu}^{o}$ and $-\operatorname{Im}%
\Sigma^{\mathrm{R}}\left(  \varepsilon_{\nu}^{o}\right)  /\hbar=1/(2\tau
_{\mathrm{SE}}^{{}})$ accounts for their\ \textquotedblleft decay
rate\textquotedblright\ into collective SE eigenstates in agreement with a
self-consistent Fermi Golden Rule (FGR) \cite{SC-FGR}, i.e. the evolution with
$\mathbf{H}_{\mathrm{eff.}}$ is non-unitary.

The complete dynamics will be obtained resorting to the Keldysh formalism
\cite{Keldysh,Keldysh2}. This allows the evaluation of the relevant
density-density correlations within a norm conserving scheme \cite{Keldysh2}:%
\begin{multline}
\mathbf{G}^{<}\left(  t_{2},t_{1}\right)  =\hbar^{2}\mathbf{G}^{\mathrm{R}%
}\left(  t_{2},0\right)  \mathbf{G}^{<}\left(  0,0\right)  \mathbf{G}%
^{\mathrm{A}}\left(  0,t_{1}\right)  +\\
\int_{0}^{t_{2}}\int_{0}^{t_{1}}\mathrm{d}t_{k}\mathrm{d}t_{l}\mathbf{G}%
^{\mathrm{R}}\left(  t_{2},t_{k}\right)  \mathbf{\Sigma}^{<}\left(
t_{k},t_{l}\right)  \mathbf{G}^{\mathrm{A}}\left(  t_{l},t_{1}\right)  .
\label{Danielewicz_evol}%
\end{multline}
The first term stands for the \textquotedblleft coherent\textquotedblright%
\ evolution while second term contains \textquotedblleft incoherent
reinjections\textquotedblright, described by the injection self-energy,
$\mathbf{\Sigma}^{<},$ that compensates any leak from the coherent evolution
\cite{GLBE2}. Solving Eq. (\ref{Danielewicz_evol}) requires the particle
(hole) injection and retarded self-energies, $\Sigma^{<(>)}\left(  t_{1}%
,t_{2}\right)  $ and $\Sigma^{\mathrm{R}}\left(  t_{1},t_{2}\right)
=\theta\left(  t_{1},t_{2}\right)  [\Sigma^{>}\left(  t_{2},t_{1}\right)
-\Sigma^{<}\left(  t_{2},t_{1}\right)  ]$. For this, we use a perturbative
expansion on $\widehat{\mathcal{H}}_{\mathrm{SE}}$. The first order gives the
standard Hartree-Fock energy corrections which, being real, do not contribute
to $\Sigma^{<}$. The second order terms \cite{Keldysh2}, sketched in Fig.
\ref{Fig_feynmann_ising} \textit{b)}, contribute to\ $\ \Sigma^{\gtrless}$,
and in the space representation:
\begin{equation}
\frac{\Sigma_{ij}^{\lessgtr}\left(  t_{k},t_{l}\right)  }{\hbar^{2}%
}=\left\vert U_{is}^{{}}\right\vert ^{2}G_{ss}^{\lessgtr}\left(  t_{k}%
,t_{l}\right)  G_{ss}^{\gtrless}\left(  t_{l},t_{k}\right)  G_{ii}^{\lessgtr
}\left(  t_{k},t_{l}\right)  ~\delta_{ij}, \label{Sigma_Feynmann}%
\end{equation}
where $\left(  i,s\right)  $ $\in$ $\left\{  \left(  A,\mathrm{L}\right)
,\left(  B,\mathrm{R}\right)  \right\}  $ and hence $s$ stands for the surface
site of the environment. The net interaction between an electron in the system
and one in the reservoir is $U_{is}$ =$-2U_{is}^{(\mathrm{dir.})}%
+U_{is}^{(\mathrm{exch.})},$where the direct term contributes with a fermion
loop with an extra spin summation. Notice that self-energy diagrams shown in
Fig. \ref{Fig_feynmann_ising} \textit{b) } describe an electron exciting an
electron-hole pair in the environment and later destroying it. The evaluation
of these processes requires accounting for the different time scales for the
propagation of excitations in the system and reservoirs. We resort to
time-energy variables \cite{GLBE2}: $t_{i}=\frac{t_{k}+t_{l}}{2}$, the
physical time, and $\delta t_{i}=t_{k}-t_{l},$ which characterizes the quantum
correlations. The integrand in Eq. (\ref{Danielewicz_evol}), when $t_{2}%
=t_{1}=t$ becomes $\mathbf{G}^{\mathrm{R}}\left(  t,t_{i}+\delta
t_{i}/2\right)  \mathbf{\Sigma}^{<}\left(  \delta t_{i},t_{i}\right)
\mathbf{G}^{\mathrm{A}}\left(  t_{i}-\delta t_{i}/2,t\right)  $. Since $\delta
t_{i}$ is related to the energy $\varepsilon$ through a \ Fourier transform
(FT), in equilibrium,%
\begin{align}
G_{s,s}^{<}\left(  \varepsilon,t_{i}\right)   &  =\mathrm{i}2\pi~N_{s}\left(
\varepsilon\right)  ~\mathrm{f}_{s}\left(  \varepsilon,t_{i}\right)
,\label{G<(E,t)}\\
G_{s,s}^{>}\left(  \varepsilon,t_{i}\right)   &  =-\mathrm{i}2\pi~N_{s}\left(
\varepsilon\right)  ~\left[  1-\mathrm{f}_{s}\left(  \varepsilon,t_{i}\right)
\right]  .\nonumber
\end{align}
Here $N_{s}\left(  \varepsilon\right)  $ is the local density of states
(LDoS)\ at the surface of the reservoir and $\mathrm{f}_{s}\left(
\varepsilon,t_{i}\right)  =\frac{1}{2}$ is the occupation factor in the high
temperature limit ($k_{B}T\gg V_{s}$). Replacing the LDoS \cite{RevMex}
$N_{s}\left(  \varepsilon\right)  =1/\left(  \pi V_{s}\right)  \sqrt{1-\left(
\frac{\varepsilon}{2V_{s}}\right)  ^{2}}$in Eq. (\ref{G<(E,t)}), and doing the
FT \cite{SC-FGR} one gets $G_{s,s}^{\lessgtr}\left(  \delta t_{i}%
,t_{i}\right)  =\pm\mathrm{i}\frac{1}{V_{s}}\frac{J_{1}(\frac{2V_{s}}{\hbar
}\delta t_{i})}{\delta t_{i}}\frac{1}{2},$where $J_{1}$ is the Bessel function
of first order. Replacing in Eq. (\ref{Sigma_Feynmann})
\begin{equation}
\frac{\Sigma_{ij}^{\lessgtr}\left(  \delta t_{i},t_{i}\right)  }{\hbar^{2}%
}=U_{is}^{2}\left[  \frac{J_{1}(\frac{2V_{s}}{\hbar}\delta t_{i})}%
{2V_{s}\delta t_{i}}\right]  ^{2}G_{ii}^{\lessgtr}\left(  \delta t_{i}%
,t_{i}\right)  ~\delta_{ij}. \label{Sigma<}%
\end{equation}
$\left\vert J_{1}\left[  2V_{s}\delta t_{i}/\hbar\right]  /\left(  V_{s}\delta
t_{i}\right)  \right\vert ^{2}$ decays in a time scale $\hbar/V_{s}$ which, in
the wide-band (WB) limit $V_{s}\gg V_{AB}$, is much shorter than $\hbar
/V_{AB}$, the time scale of $G_{ii}^{\lessgtr}(\delta t_{i},t_{i})$. Hence,
the main contribution to the integral on $\delta t_{i}$ in Eq.
(\ref{Danielewicz_evol}) is obtained replacing $G_{ii}^{\lessgtr}\left(
\delta t_{i},t_{i}\right)  $ by $G_{ii}^{\lessgtr}\left(  0,t_{i}\right)  $.
The same consideration holds for $\mathbf{G}^{\mathrm{R}}\left(
t,t_{i}+\delta t_{i}/2\right)  \ $ and $\mathbf{G}^{\mathrm{A}}\left(
t_{i}-\delta t_{i}/2,t\right)  $ which are replaced by $\mathbf{G}%
^{\mathrm{R}}\left(  t,t_{i}\right)  $ and\textbf{ }$\mathbf{G}^{\mathrm{A}%
}\left(  t_{i},t\right)  $. Then, the dependence on $\delta t_{i}$ enters only
through $\Sigma_{ij}^{\lessgtr}\left(  \delta t_{i},t_{i}\right)  $ yielding
\begin{align}
\Sigma_{ij}^{\lessgtr}\left(  t_{i}\right)   &  =\int\Sigma_{ij}^{\lessgtr
}\left(  \delta t_{i},t_{i}\right)  \mathrm{d}\delta t_{i}\\
&  =\frac{8}{6\pi}\frac{\left\vert U_{is}\right\vert ^{2}}{V_{s}}\hbar
~G_{ii}^{\lessgtr}\left(  t_{i}\right)  ~\delta_{ij}.
\label{sigma-Keldysh-local}%
\end{align}
Hence, in the WB limit, the Keldysh self-energy of Eq. (\ref{Sigma_Feynmann})
becomes local in space and time and no further structure from higher order
terms is admisible. This is represented as a collapse of sucessive pairs of
black dots in Fig. (\ref{Fig_feynmann_ising}) \textit{b) }into a single point.
The expansion represented in Fig. \ref{Fig_feynmann_ising} \textit{c)} and
that of Fig. \ref{Fig_feynmann_ising} \textit{d) }become exact in this limit.

We assume $E_{A}=E_{B}=E_{\mathrm{L}}=E_{\mathrm{R}}=0$ and the symmetry
condition $\left\vert U_{A\mathrm{L}}^{{}}\right\vert ^{2}/V_{\mathrm{L}%
}=\left\vert U_{B\mathrm{R}}\right\vert ^{2}/V_{\mathrm{R}}$. From Eq.
(\ref{sigma-Keldysh-local}) we obtain the decay rates
\begin{align}
\frac{1}{\tau_{\mathrm{SE}}}  &  \equiv\tfrac{2}{\hbar}\Gamma_{\mathrm{SE}%
}\equiv\tfrac{-2}{\hbar}\operatorname{Im}\Sigma_{ii}^{R}\equiv\tfrac
{\mathrm{i}}{\hbar}\left(  \Sigma_{ii}^{A}-\Sigma_{ii}^{R}\right)
\label{decay-rate}\\
&  =\tfrac{2\pi}{\hbar}\left\vert U_{is}^{{}}\right\vert ^{2}\tfrac{2}%
{3\pi^{2}V_{s}},\nonumber
\end{align}
coinciding with the FGR. The WB limit can be relaxed because the FGR holds
when time $t$ is in the range \cite{SC-FGR} $t_{S}<t<t_{R}\simeq\alpha
\hbar/\Gamma_{SE}\ln(V_{s}/\Gamma_{SE}),$ where $\alpha$ depends on the van
Hove singularities of the spectral density $\mathcal{J}_{s}(\varepsilon)=\int
N_{s}(\varepsilon-\varepsilon^{\prime})N_{s}(\varepsilon^{\prime}%
)\mathrm{d}\varepsilon^{\prime}$. Here, $t_{S}=\hbar\mathcal{J}_{s}%
(0)\simeq\hbar/V_{s}$ is the survival time of an electron-hole excitation at
the surface site and $t_{R}$ characterizes the transition to a power law decay
arrising from memory effects. Hence, as long as $\Gamma_{SE},V_{AB}\ll V_{s},$
the FGR is valid for times much longer than $\hbar/\Gamma_{SE}$. Under these
conditions, $\mathbf{H}_{\mathrm{eff.}}$ does not depend on $\varepsilon$ and
the propagator has a simple dependence on $t$ as $\mathbf{G}^{\mathrm{R}%
}\left(  t\right)  =\mathbf{G}^{0\mathrm{R}}\left(  t\right)  e^{-\Gamma
_{\mathrm{SE}}t/\hbar},$ where, $G_{AA}^{0\mathrm{R}}(t)=G_{BB}^{0\mathrm{R}%
}(t)=\frac{\mathrm{i}}{\hbar}\cos\left(  \frac{\omega_{0}}{2}t\right)  $ and
$G_{AB}^{0\mathrm{R}}(t)=G_{BA}^{0\mathrm{R}}(t)=\frac{\mathrm{i}}{\hbar}%
\sin\left(  \frac{\omega_{0}}{2}t\right)  $. Eq. (\ref{Danielewicz_evol}) becomes,%

\begin{multline}
\mathbf{G}^{<}\left(  t,t\right)  =\hbar^{2}\mathbf{G}^{0\mathrm{R}}\left(
t\right)  \mathbf{G}^{<}\left(  0,0\right)  \mathbf{G}^{0\mathrm{A}}\left(
-t\right)  e^{-t/\tau_{\mathrm{SE}}}+\label{danielewicz2}\\
\int_{0}^{t}\mathrm{d}t_{i}\mathbf{G}^{0\mathrm{R}}\left(  t-t_{i}\right)
\mathbf{\Sigma}^{<}\left(  t_{i}\right)  \mathbf{G}^{0\mathrm{A}}\left(
t_{i}-t\right)  e^{-\left(  t-t_{i}\right)  /\tau_{\mathrm{SE}}},
\end{multline}
a generalized Landauer-B\"{u}ttiker equation \cite{GLBE1,GLBE2}. The initial
condition has the state $A$ occupied: $\frac{\hbar}{\mathrm{i}}G_{ij}%
^{<}\left(  0,0\right)  =\delta_{iA}\delta_{Aj}$. Replacing Eq.
(\ref{sigma-Keldysh-local}) into Eq. (\ref{danielewicz2}),\ identifying the
interaction rate of Eq. (\ref{decay-rate}) we get two coupled equations for
$G_{AA}^{<}$ and $G_{BB}^{<}$%
\begin{gather}
\tfrac{\hbar}{\mathrm{i}}G_{%
\genfrac{}{}{0pt}{}{AA}{(BB)}%
}^{<}\left(  t,t\right)  =\hbar^{2}\left\vert G_{%
\genfrac{}{}{0pt}{}{AA}{(BA)}%
}^{0\mathrm{R}}\left(  t\right)  \right\vert ^{2}e^{-t/\tau_{\mathrm{SE}}%
}\nonumber\\
+\int\hbar^{2}\left\vert G_{%
\genfrac{}{}{0pt}{}{AA}{(BA)}%
}^{0\mathrm{R}}\left(  t-t_{i}\right)  \right\vert ^{2}e^{-(t-t_{i}%
)/\tau_{\mathrm{SE}}}\frac{\mathrm{d}t_{i}}{\tau_{\mathrm{SE}}}\left[
\tfrac{\hbar}{\mathrm{i}}G_{%
\genfrac{}{}{0pt}{1}{AA}{{}}%
}^{<}\left(  t_{i}\right)  \right]  \nonumber\\
+\int\hbar^{2}\left\vert G_{%
\genfrac{}{}{0pt}{}{AB}{(BB)}%
}^{0\mathrm{R}}\left(  t-t_{i}\right)  \right\vert ^{2}e^{-(t-t_{i}%
)/\tau_{\mathrm{SE}}}\frac{\mathrm{d}t_{i}}{\tau_{\mathrm{SE}}}\left[
\tfrac{\hbar}{\mathrm{i}}G_{%
\genfrac{}{}{0pt}{1}{BB}{{}}%
}^{<}\left(  t_{i}\right)  \right]  .\label{GLBE-2x2}%
\end{gather}
The first term is the probability that a particle initially at site $A$ is
found in site $A$ (or $B$) at time $t$ having survived the interactions with
the environment. The second and third terms describe particles whose last
interaction with the environment, at time $t_{i}$, occured at site $A$ and $B$
respectively. The Laplace transform yields:%
\begin{equation}
\tfrac{\hbar}{\mathrm{i}}G_{AA}^{<}\left(  t,t\right)  =\tfrac{1}{2}%
+a_{0}~\cos\left[  \left(  \omega+\mathrm{i}\eta\right)  t-\phi\right]
e^{-t/\left(  2~\tau_{\mathrm{SE}}^{{}}\right)  },\label{G11}%
\end{equation}
where $a_{0}^{2}=\left(  4\omega^{2}\tau_{\mathrm{SE}}^{2}+1\right)  /\left(
16\omega^{2}\tau_{\mathrm{SE}}^{2}\right)  ;$ $\phi=\arctan\left[  1/\left(
2\omega\tau_{\mathrm{SE}}^{{}}\right)  \right]  $ and
\begin{align}
\omega &  =\left\{
\begin{array}
[c]{cc}%
\omega_{0}\sqrt{1-\left(  2\omega_{0}\tau_{\mathrm{SE}}^{{}}\right)  ^{-2}} &
~~~~\omega_{0}>\frac{1}{2\tau_{\mathrm{SE}}^{{}}}\\
0 & ~~~~\omega_{0}\leq\frac{1}{2\tau_{\mathrm{SE}}^{{}}}%
\end{array}
\right.  ,\label{frequency}\\
\eta &  =\left\{
\begin{array}
[c]{cc}%
0 & ~~~~\omega_{0}>\frac{1}{2\tau_{\mathrm{SE}}^{{}}}\\
\omega_{0}\sqrt{\left(  2\omega_{0}\tau_{\mathrm{SE}}^{{}}\right)  ^{-2}-1} &
~~~~\omega_{0}\leq\frac{1}{2\tau_{\mathrm{SE}}^{{}}}%
\end{array}
\right.  .\label{relaxation_w}%
\end{align}
Noticeably, in the first term of Eq. (\ref{danielewicz2}) the environment,
though giving the exponential decay, does not affect the frequency.
Modification of $\omega$ requires the dynamical feedback.

The effect of lateral chains on the two state system can produce observables
with non-linear dependences on $H_{\mathrm{SE}}$ which could account for a
cross over among the limiting dynamical regimes. However, we find a
\textit{non-analyticity} in these functions enabled by the infinite degrees of
freedom of the environment \cite{sachdev} (i.e. the thermodynamic limit).
Here, they are incorporated through the imaginary part of the self-energy,
$\hbar/\tau_{\mathrm{SE}},$ i.e. the FGR. \ Hence, the non-analyticity of
$\omega$ and $\tau_{\phi}$ on the control parameter $\omega_{0}\tau
_{\mathrm{SE}}$ at the critical value, indicates a switch between two
dynamical regimes which we call a Quantum Dynamical Phase Transition.

In the \textit{swapping phase} the observed frequency $\omega$ is finite.
According to Eq. (\ref{frequency}), if $\omega_{0}\tau_{\mathrm{SE}}\gg1$ \ it
coincides with $\omega_{0},$ indicating a weakly perturbed evolution. As one
approaches the critical value $\omega_{0}\tau_{\mathrm{SE}}$=$\frac{1}{2}$,
$\omega$ decreases vanishing at the critical point. Beyond that value lies the
\textit{Zeno phase} where the swapping freezes ($\omega=0).$

\begin{figure}
[h]
\begin{center}
\includegraphics[
height=2.4215in,
width=1.9588in
]%
{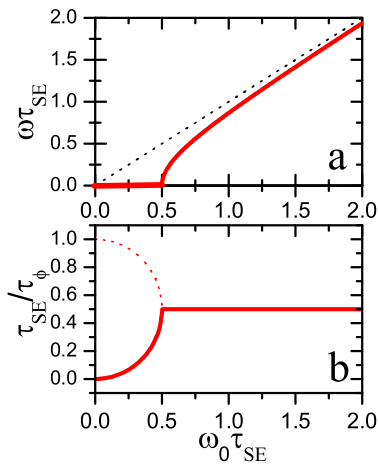}%
\caption{(Color online) a) Swapping frequency as a function of the control
parameter $\omega_{0}\tau_{SE}.$ Departure from the dashed line is a
consequence of the environment. b) Decoherence rate as a function of
$\omega_{0}\tau_{SE}.$ At $\omega_{0}\tau_{SE}=1/2$ there is a bifurcation
into two decay modes.}%
\label{fig_wyR}%
\end{center}
\end{figure}

The \textquotedblleft decoherence\textquotedblright\ rate observed from the
attenuation of the oscillation is:%
\begin{equation}
1/\tau_{\phi}=1/\left(  2\tau_{\mathrm{SE}}^{{}}\right)  ~~~~~~\text{for}%
~~~\omega_{0}\geq1/\left(  2\tau_{\mathrm{SE}}\right)  .
\end{equation}
The dependence of the first term in Eq. (\ref{danielewicz2}) on $1/\tau
_{\mathrm{SE}}$ describes the decay of the initial state into
system-environment superpositions. This decay is instantaneously compensated
by the \textquotedblleft reinjection\textquotedblright\ term which being
\textquotedblleft in-phase\textquotedblright\ with the Rabi oscillation
ensures $1/\tau_{\phi}$ $\leq1/\tau_{\mathrm{SE}}$. Beyond the critical value,
$\omega_{0}\tau_{\mathrm{SE}}\leq1/2$, the decay rate in Eq. (\ref{G11})
\textit{bifurcates} in \textit{two} damping modes $1/\left(  2\tau
_{\mathrm{SE}}\right)  \pm\eta$. The slowest, for$~~\omega_{0}\leq\tfrac
{1}{2\tau_{\mathrm{SE}}},$ is:%
\begin{equation}
1/\tau_{\phi}=\tfrac{1}{2\tau_{\mathrm{SE}}}\left[  1-\sqrt{1-\left(
2\omega_{0}\tau_{\mathrm{SE}}^{{}}\right)  ^{2}}\right]
\genfrac{}{}{0pt}{1}{{}}{\overrightarrow{\omega_{0}^{{}}\tau_{\mathrm{SE}}%
^{{}}\rightarrow0}}%
\omega_{0}^{2}\tau_{\mathrm{SE}}^{{}}.
\end{equation}
This manifests the Quantum Zeno Effect: the stronger the interaction with the
environment, the longer the survival of the initial state. The critical
behavior of the observables $\omega$ and $1/\tau_{\phi}$ is shown in Fig.
\ref{fig_wyR} a)\textbf{ }and b).

\begin{figure}
[tb]
\begin{center}
\includegraphics[
height=3.9314in,
width=3.1704in
]%
{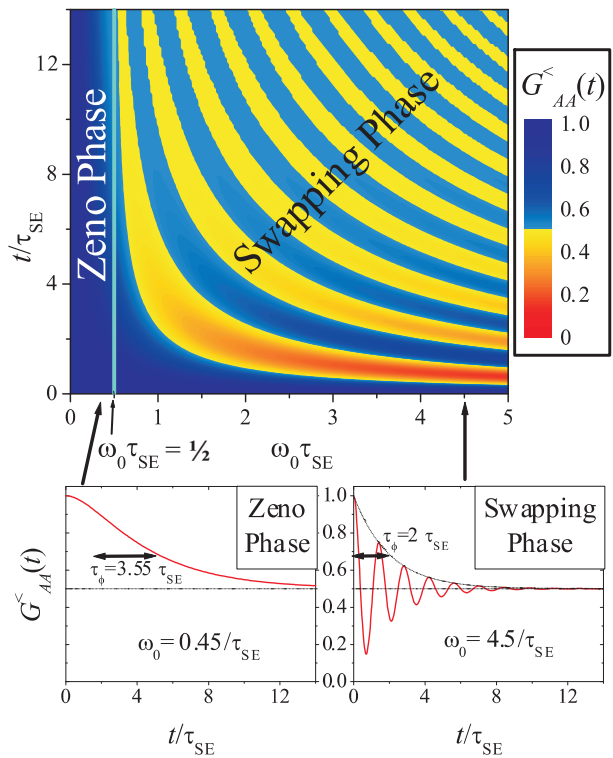}%
\caption{(Color online) Contour plot of \ the particle density function
$G_{AA}^{<}(t)$ as a function of time and $\omega_{0}$ in units of
$\tau_{\text{\textrm{SE}}}$. The vertical line indicates the critical value
where the oscillation period diverges. Panels at the bottom show the behavior
of $G_{AA}^{<}(t)$ for values of $\omega_{0}\tau_{\text{\textrm{SE}}}$ within
the Zeno phase (left panel) and the Swapping phase (right panel).}%
\label{Fig_stripes}%
\end{center}
\end{figure}

In Fig. \ref{Fig_stripes} different colors label excess or defect in the
occupation of state $A$ with respect to equilibrium. The hyperbolic stripes
show that the swapping period $T=2\pi/\omega$ diverges beyond a finite
critical value $\omega_{0}\tau_{\mathrm{SE}}=1/2$, evidencing the
\emph{dynamical phase transition}. Near the critical point $T\simeq
(T_{\mathrm{c}}/\sqrt{2})\left(  1-T_{0}/T_{0}^{\mathrm{c}}\right)  ^{-1/2},$
where $T_{0}=\frac{2\pi}{\omega_{0}}$ and $T_{0}^{\mathrm{c}}=4\pi
\tau_{\mathrm{SE}}$ is the critical natural period. The critical exponent is
$-1/2$. Typical dynamics illustrating both phases are shown at the bottom.
Both start with a quadratic decay which is beyond the FGR and results from the
phase $\phi$ in Eq. (\ref{G11}).%

Related dynamical regimes are predicted for different dissipative two level
systems as the spin-boson model initially addressed by Chakravarty and Leggett
\cite{spin-boson}. While in that model the memory effects are fundamental, our
system allows one to reach the strong coupling limit in a case where they are
not relevant. Hence, both, $\omega$ and $\tau_{\phi},$ get simple but
non-analytic expressions in terms of the control parameter manifesting the
critical aspects of the transition to the Zeno phase. This transition has been
observed in NMR as an abrupt drop in the frequency and relaxation rate
\cite{JCP98, JCP06}. The present theory also maps to these spin systems
\cite{JCP06}.

In summary, a microscopic model for an electron in a two state system coupled
to an environment through a many-body interaction shows that by sweeping
$\omega_{0}\tau_{\mathrm{SE}}$ below the critical value $\frac{1}{2}$ the
oscillatory dynamics freezes. While a crossover to a Zeno regime is expected
for strong SE interaction, a non-analytic point separating a finite parametric
region of dynamical freeze from a swapping phase defines a \textit{quantum
dynamical phase transition}.


\begin{thebibliography}{99}                                                                                               %


\bibitem {Zurek2003}W. H. Zurek, Rev. Mod. Phys. \textbf{75}, 715 (2003).

\bibitem {PhysicaA}H. M. Pastawski \textit{et al}., Physica A \textbf{283},
166 (2000).

\bibitem {JalPas}R. A. Jalabert and H. M. Pastawski, Phys. Rev. Lett\textit{.}
\textbf{86}, 2490 (2001).

\bibitem {Beenakker}Ph. Jacquod, P. G. Silvestrov and C. W. J. Beenakker,
Phys. Rev. E\textit{ }\textbf{64}, 055203(R) (2001).

\bibitem {CookPasJal}F. M. Cucchietti, H. M. Pastawski and R. A. Jalabert,
Phys. Rev. B \textbf{70}, 035311 (2004).

\bibitem {exp-swapp}C. J. Myatt \textit{et al.}, Nature \textbf{403}, 269 (2000).

\bibitem {Nakamura}Y. Nakamura, Yu. A. Pashkin and J. S. Tsai, Nature
\textbf{398}, 786 (1999).

\bibitem {Misra-Sudarshan}B. Misra and E. C. G. Sudarshan; J. Math. Phys.
\textbf{18}, 756 (1977).

\bibitem {Pascazio}S. Pascazio and M. Namiki, Phys. Rev. A \textbf{50} 4582 (1994).

\bibitem {Pastawski-Usaj}H. M. Pastawski and G. Usaj, Phys. Rev. B
\textbf{57}, 5017 (1998).

\bibitem {DAmato}P. R. Levstein, H. M. Pastawski and J. L. D'Amato, J. Phys.:
Condens. Matter\textit{ }\textbf{2,} 1781 (1990).

\bibitem {SC-FGR}E. Rufeil Fiori and H.M. Pastawski, \textit{Chem. Phys. Lett.
}\textbf{420}\textit{ }35 (2006).

\bibitem {Keldysh}L. V. Keldysh, \textit{ZhETF} \textbf{47}, 1515 (1964)
[\textit{Sov. Phys.-JETP} \textbf{20}, 1018 (1965)].

\bibitem {Keldysh2}P. Danielewicz, Ann. Phys. \textbf{152}, 239 (1984).

\bibitem {GLBE2}H. M. Pastawski, Phys. Rev. B \textbf{46}, 4053 (1992).

\bibitem {RevMex}H. M. Pastawski and E. Medina, Rev. Mex. de F\'{\i}sica
\textbf{47S1}, 1 (2001); cond-mat/0103219.

\bibitem {GLBE1}H. M. Pastawski, Phys. Rev. B \textbf{44}, 6329 (1991).

\bibitem {sachdev}S. Sachdev, \textit{Quantum Phase Transitions,} (Cambridge
U. P., 2001).

\bibitem {spin-boson}S. Chakravarty and A. J. Leggett, Phys. Rev. Lett.
\textbf{52}, 5 (1984); U. Weiss, \textit{Quantum Dissipative Systems} (World
Scientific, Singapore, First Edition, 1993).

\bibitem {JCP98}P. R. Levstein, G. Usaj and H. M. Pastawski, J. Chem. Phys.
\textbf{108}, 2718 (1998) (see Fig. 5 and Fig. 7).

\bibitem {JCP06}G. A. \'{A}lvarez, E. P. Danieli, P. R. Levstein and H. M.
Pastawski, accepted for its publication in J. Chem. Phys. (2006); condmat-0504347.
\end{thebibliography}
\end{document}